\documentclass[aip,amsmath,amssymb,reprint,graphicx]{revtex4-1}
\usepackage{graphicx}
\usepackage{dcolumn}
\usepackage{bm}
\usepackage{float}

\usepackage[utf8]{inputenc}
\usepackage[T1]{fontenc}
\usepackage{mathptmx}
\usepackage{siunitx}  
\usepackage{amsmath}
\usepackage{subfigure}
\usepackage[hidelinks]{hyperref}
\usepackage[all]{hypcap}
\usepackage[noabbrev, capitalise]{cleveref}
\usepackage{amssymb}

\usepackage{natbib}

\draft

\begin{document}

\preprint{AIP/123-QED}

\title{Nanomechanical resonators fabricated by atomic layer deposition on suspended 2D materials} 

\author{Hanqing Liu}
\affiliation{Department of Precision and Microsystems Engineering, Delft University of Technology, Mekelweg 2, 2628 CD, Delft, The Netherlands}

\author{Saravana B. Basuvalingam}
\affiliation{Department of Applied Physics, Eindhoven University of Technology, P.O. Box 513, 5600 MB,
Eindhoven, The Netherlands}

\author{Saurabh Lodha}
\email{Lodha@ee.iitb.ac.in}
\affiliation{Department of Electrical Engineering, Indian Institute of Technology Bombay, 400076, India}

\author{Ageeth A. Bol}
\affiliation{University of Michigan, Department of Chemistry, 930 N. University Ave., Ann Arbor, MI, 48109, USA}

\author{Herre S. J. van der Zant}
\affiliation{Kavli Institute of Nanoscience, Delft University of Technology, Lorentzweg 1, 2628 CJ, Delft, The Netherlands}

\author{Peter G. Steeneken}
\affiliation{Department of Precision and Microsystems Engineering, Delft University of Technology, Mekelweg 2, 2628 CD, Delft, The Netherlands}
\affiliation{Kavli Institute of Nanoscience, Delft University of Technology, Lorentzweg 1, 2628 CJ, Delft, The Netherlands}

\author{G. J. Verbiest}%
\email{G.J.Verbiest@tudelft.nl}
\affiliation{Department of Precision and Microsystems Engineering, Delft University of Technology, Mekelweg 2, 2628 CD, Delft, The Netherlands}

\begin{abstract}
Atomic layer deposition (ALD), a layer-by-layer controlled method to synthesize ultrathin materials, provides various merits over other techniques such as precise thickness control, large area scalability and excellent conformality. Here we demonstrate the possibility of using ALD growth on top of suspended 2D materials to fabricate nanomechanical resonators. We fabricate ALD nanomechanical resonators consisting of a graphene/MoS$_2$ heterostructure. Using AFM indentation and optothermal drive, we measure their mechanical properties including Young's modulus, resonance frequency and quality factor, showing similar values as their exfoliated and chemical vapor deposited counterparts. We also demonstrate the fabrication of nanomechanical resonators by exfoliating an ALD grown NbS$_2$ layer. This study exemplifies the potential of ALD techniques to produce high-quality suspended nanomechanical membranes, providing a promising route towards high-volume fabrication of future multilayer nanodevices and nanoelectromechanical systems.

\end{abstract}

\maketitle

The properties of 2D materials, in particular their ultralow weight and ultrahigh mechanical flexibility, provides them with an excellent sensitivity to external forces \cite{barton2011high,zhu2022frequency,lee2013high}. Hence, resonators from 2D materials have become a popular choice for the next generation of nanoelectromechanical systems (NEMS) \cite{steeneken2021dynamics,lemme2020nanoelectromechanical}. Recently, there is surge towards stacking different 2D materials into heterostructures often exhibiting better sensing properties. Such heterostructures are used for tunable resonators and oscillators \cite{ye2021ultrawide}, and can potentially lead to better sensors in microphone and pressure sensing applications \cite{lemme2020nanoelectromechanical}.

To achieve high-performance nanomechanical resonators, clean interfaces between different 2D materials are important \cite{mackus2014use}. Therefore, bottom-up synthesis methods were developed, of which chemical vapor deposition (CVD) is the most attractive due to its large-scale and high-quality growth. The main shortcoming of CVD, however, is the difficulty to accurately control the thickness and morphology of grown 2D materials. Atomic layer deposition (ALD), a vapor phase thin film deposition technique based on self-limiting surface reactions, inherently yields atomic-scale thickness control, excellent uniformity, and conformality \cite{sharma2018low}. ALD processes exists for a large variety of materials ranging from pure elements to metal oxides and chalcogenides \cite{vos2016atomic}. In terms of 2D materials, ALD was applied to fabricate 2D-based field effect transistors, p-n diode devices, solar cells and photodetectors, displaying high electrical and optical uniformities \cite{hao2018atomic}. Since layer thickness control is essential for realizing uniform mechanical properties, it is of interest to explore the potential of ALD materials for nanomechancial resonators.

In this letter, we show two types of nanomechancial resonators fabricated using ALD: one consists of a heterostructure made from exfoliated graphene (bottom layer) and ALD MoS$_2$ (top layer) and the other is ALD NbS$_2$. We use atomic force microscope (AFM) indentation to determine their Young's moduli and use an optomechanical method to study their resonance frequency and corresponding quality factor in vacuum conditions. The extracted parameters from our measurements agree well with literature values for 2D exfoliated or CVD resonators. Furthermore, by fitting a relation between the quality factors before and after ALD, we verify a low-level dissipation induced by ALD MoS$_2$. Our work indicates the potential of ALD fabrication techniques for realizing multilayer nanomechanical membranes and resonators with enhanced functionality and thickness control.

\begin{figure*}
	\centering
	\includegraphics[width=16cm]{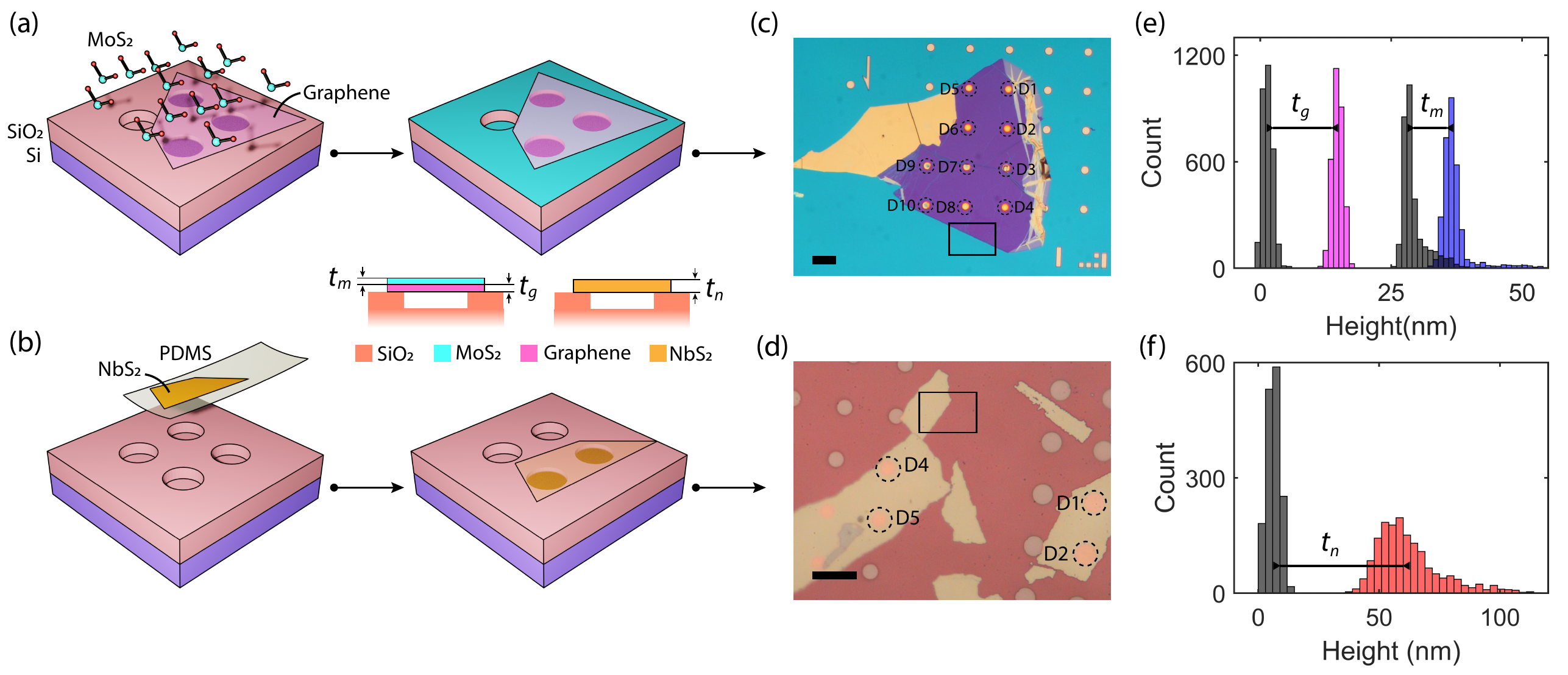}
	\caption{Fabrication of the ALD nanomechanical resonators measured in this work. (a) Fabrication process of exfoliated graphene/ALD MoS$_2$ heterostructure resonators. MoS$_2$ is directly deposited onto the suspended graphene membrane. (b) Fabrication process of ALD NbS$_2$ resonators. ALD NbS$_2$ flake is exfoliated from glassy carbon using PDMS, and then transferred on top of a prepatterned Si/SiO$_2$ substrate with circular cavities. Inserts between (a) and (b) are the cross-section views of the heterostructure and NbS$_2$ resonators, respectively. (c) and (d) Optical images of our fabricated heterostructure and NbS$_2$ devices, respectively, where the black scale bars are \SI{20}{\micro m}. Note that the heterostructure devices D3 and D9 were broken during deposition. (e) Height histograms of SiO$_2$/Si substrate (black), graphene (pink) and MoS$_2$ (blue) measured by AFM scanning, allow us to extract the thicknesses $t_g=$~\SI{13.3}{nm}, $t_m=$~\SI{7.8}{nm}, and $t_n=$~\SI{56.1}{nm}. The scanning areas for fixing $t_g$ and $t_n$ are the black frames in (c) and (d), respectively, while that for fixing $t_m$ is the scratch on MoS$_2$ layer after deposition. We observe that the thickness distribution of ALD MoS2 is satisfyingly as uniform as the exfoliated graphene, while the surface of thick ALD NbS$_2$ is rugged.         
	}
	\label{fig: fabrication}
\end{figure*} 


The ALD layers are deposited (see Figs.~\ref{fig: fabrication}a and b) by plasma-enhanced atomic layer deposition (PE-ALD) technique using an Oxford Instruments Plasma Technology FlexAL ALD reactor. The base pressure of the system is 10$^{-6}$~\si{Torr}. The metal-organic precursors bis(tert-butylimido)-bis(dimethylamido) molybdenum (STREM Chemical, Inc., 98$\%$) and (tert-butylimido)-tris-(diethylamino)-niobium (STREM Chemical, Inc., 98$\%$) are used for MoO$_x$ and NbO$_x$ growth, respectively \cite{vos2016atomic,basuvalingam2018comparison}. The Mo and Nb precursors are kept in stainless steel bubblers at 50~\si{^{\circ} C} and 65~\si{^{\circ} C}, respectively and are bubbled using Ar as the carrier gas. In both the processes, O$_2$ plasma is used as the coreactant. The MoO$_x$ and NbO$_x$ films are deposited at 100~\si{^{\circ} C} and 150~\si{^{\circ} C}, respectively. 

Both the MoS$_2$ and NbS$_2$ films are synthesized by a two step approach. As the first step, metal oxide (MoO$_x$ or NbO$_x$) film is deposited by PE-ALD technique. Next, the metal oxide film is sulfurized at 900~\si{^{\circ} C} in H$_2$S environment (10$\%$ H$_2$S and 90$\%$ Ar) to form metal sulfide film (MoS$_2$ or NbS$_2$). As shown in Figs.~\ref{fig: fabrication}a and \ref{fig: fabrication}c, the MoS$_2$ film is synthesized by PE-ALD on top of suspended graphene drums, resulting in 10 resonators with a radius $r=$ \SI{4}{\micro\meter}. Note that device D3 and D9 broke (buckled, Fig.~\ref{fig: fabrication}c) during fabrication and will not be considered further. On the other hand, NbS$_2$ film is synthesized by growing NbO$_x$ on glassy carbon followed by sulfurization at 900~\si{^{\circ} C}. Then, we fabricate the NbS$_2$ resonators by transferring the NbS$_2$ films from glassy carbon substrate over circular cavities in a SiO$_2$/Si substrate to form suspended drums using the Scotch tape method \cite{lemme2020nanoelectromechanical} (see Fig.~\ref{fig: fabrication}b). The cavities have a depth of \SI{285}{nm} and were fabricated by reactive ion etching \cite{steeneken2021dynamics}. The fabricated NbS$_2$ resonators, shown in Fig.~\ref{fig: fabrication}d, have a radius of $r=$~\SI{4}{\micro\meter} (devices D1 and D2) or $r=$~\SI{3}{\micro\meter} (devices D4 and D5). We use AFM (tapping mode) to scan the surface of our fabricated samples, to determine the thickness of the 2D materials. By calculating the height difference between the membranes and substrate (see the statistics in Figs.~\ref{fig: fabrication}e and \ref{fig: fabrication}f), we extract the mean thickness of the graphene $t_g=$~\SI{13.3}{nm} (40 layers), MoS$_2$ $t_m=$~\SI{7.8}{nm} (13 layers) and NbS$_2$ $t_n=$ \SI{56.1}{nm} (92 layers), respectively. The total thickness of the heterostructure is thus $t_h=t_m+t_g=$ \SI{21.1}{nm}.  

\begin{figure}
	\centering
	\includegraphics[height=7.5cm]{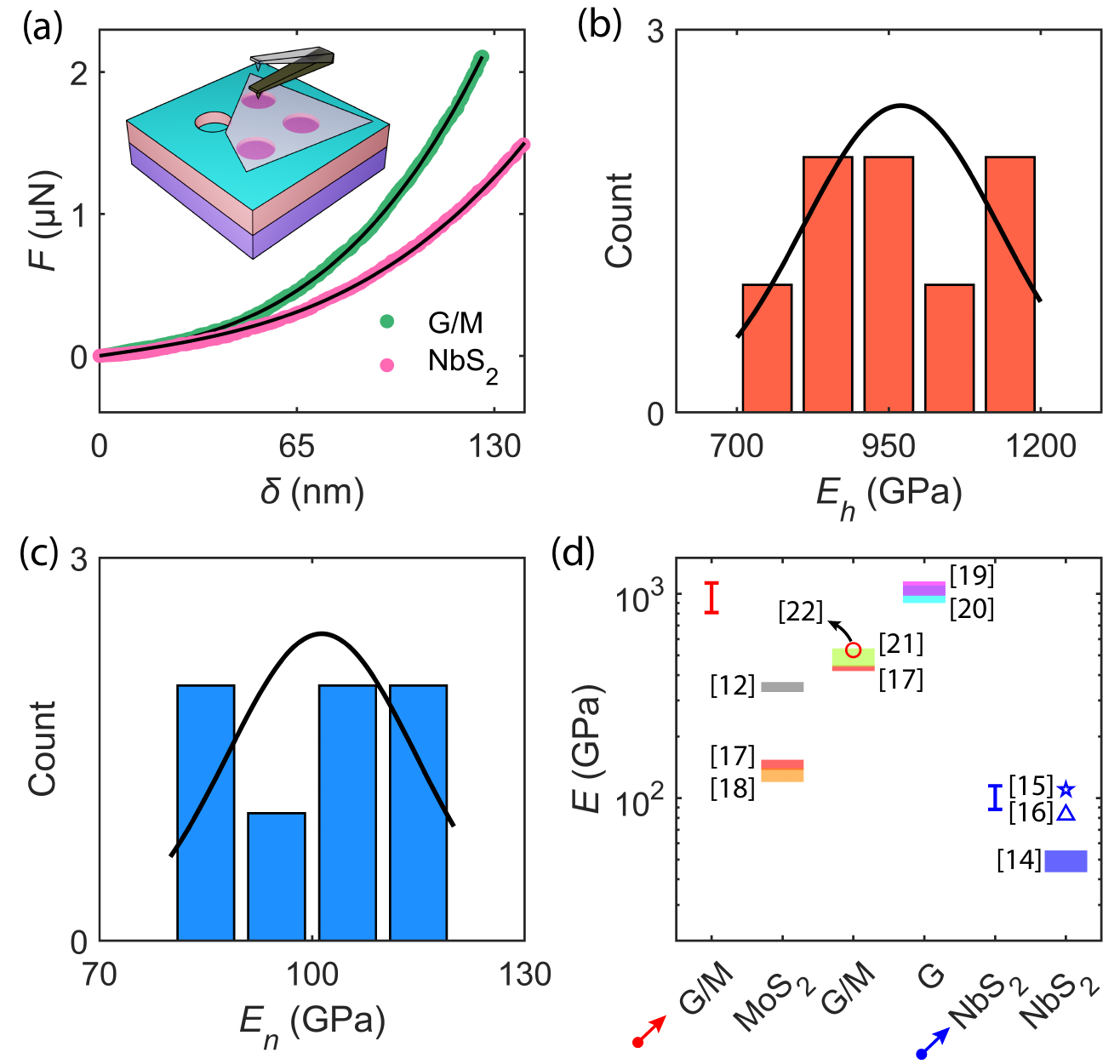}
	\caption{Static characterization of the fabricated ALD nanomechanical resonators. (a) Typical force-deflection curve obtained by AFM indentation onto the suspended part of heterostructure (green points) and NbS$_2$ (pink points) resonators. The solid lines show fits using the model from Eq.~\ref{ind}, gives the pretension and Young's modulus of the membranes. Insert: schematic of AFM indentation. (b) and (c) Histograms of $E_h$ (heterostructure) and $E_n$ (NbS$_2$), respectively, extracted from the fits to Eq. \ref{ind}. Statistics over all devices give the mean values of $Eh=$~\SI{952}{GPa} and $En=$~\SI{101}{GPa}. (d) Comparison of Young's modulus between this work (errorbars, marked with arrows in $x$-axis) and previous studies (bars and points). The numbers in brackets give the citations we select in (d).} 
	\label{fig: static}
\end{figure}

After fabrication, we determine the Young's modulus of the ALD devices by indenting with an AFM (contact mode) cantilever at the centre of the suspended area (see Fig.~\ref{fig: static}a, insert). Following literature \cite{castellanos2012elastic}, the applied vertical force $F$ versus membrane deflection $\delta$ for a circular membrane (composed of single material), as depicted in Fig.~\ref{fig: static}a, is given by
\begin{equation}
    \label{ind}
    F = \left(\frac{16\pi D}{r^2}\right) \delta+n_0\pi\delta+Etq^3 \left(\frac{\delta^3}{r^2}\right),
\end{equation}
where $D=\gamma Et^3/(12(1-\nu^2))$ is bending rigidity, $E$ is Young’s modulus, $\nu$ is the Poisson ratio, $n_0$ is pretension, $q=1/(1.05-0.15\nu-0.16\nu^2)$, and $\gamma$ is a factor that quantifies the effect of interlayer shear interactions on $D$ in multilayer 2D materials \cite{wang2019bending}. The first two terms in Eq.~\ref{ind} scale linearly with $\delta$ ($F\sim\delta$) and are set by $D$ and $n_0$; while the third cubic term ($F\sim\delta^3$) is due to the geometric nonlinearity of the membrane, which lead to an increase in the in-plane stress that depends on its Young's modulus $E$. Note that Eq.~\ref{ind} is suitable for NbS$_2$, while for heterostructures, it contains contributions from graphene and MoS$_2$ layers (see Eq.~\ref{eq:S2}). We use the bulk Poisson ratios $\nu_g=0.165$, $\nu_m=0.25$ and $\nu_n=0.28$ of graphene, MoS$_2$ and NbS$_2$, respectively, in further analysis. In addition, considering the measured layer numbers of graphene, MoS$_2$ and NbS$_2$ membranes, we use the factors $\gamma_g=0.1$ and $\gamma_m=0.4$ from literature \cite{wang2019bending} and assume $\gamma_n= \gamma_m$.

We extract $E_h$ and $E_n$ by the fitting the measured $F$ versus $\delta$ with Eq.~\ref{eq:S2} and Eq.~\ref{ind}, respectively, which nicely describe the experimental data for the NbS$_2$ device D1 (pink points) and the heterostructure device D2 (green points) as shown in Fig.~\ref{fig: static}a. Figs.~\ref{fig: static}b and \ref{fig: static}c show the extracted statistics of effective Young's moduli for heterostructure devices ($E_h$) and Young's moduli ($E_n$) for NbS$_2$ devices, giving the mean values of $E_h=952\pm$~\SI{161}{GPa} and $E_n=101\pm$~\SI{13}{GPa}, respectively. In Fig.~\ref{fig: static}d, we compare $E_h$ and $E_n$ with values reported in the literature: $E_n$ shows a good agreement with the reported values of $75\pm$ \SI{35}{GPa} \cite{sheraz2021high,sun2021first,sun2018bipolar}; $E_h$ is between the reported values for MoS$_2$ 250~$\pm$~\SI{120}{GPa} and graphene membranes 1025~$\pm$ \SI{125}{GPa} \cite{elder2015stacking,castellanos2012elastic,bertolazzi2011stretching,lopez2015increasing,lee2008measurement}, but higher than the reported values for similar fully exfoliated heterostructures 461~$\pm$~\SI{43}{GPa} \cite{elder2015stacking,ye2017atomic,liu2014elastic}. The larger $E_h$ might be caused by the stronger interlayer adhesion or the larger intrinsic Young's modulus of ALD MoS$_2$. The standard deviations in extracted Young's moduli, $\pm$\SI{13}{GPa} and $\pm$\SI{161}{GPa} for NbS$_2$ and heterostructure resonators, respectively, are comparable to the ones reported in literature for exfoliated materials. This illustrates the high homogeneity of ALD materials. 

In addition to the Young's modulus, we also extract the pretension $n_0$ for each device. Supplementary tables \ref{table: database1} and \ref{table: database2} show a complete overview of the obtained parameters from the fitting to Eq. \ref{ind}. The extracted $n_0$ ranges from 0.45 to \SI{1.55}{N/m} for all heterostructure and NbS$_2$ resonators, which are similar to values reported in the literature for resonators made by exfoliation and CVD \cite{ye2017atomic,steeneken2021dynamics,vsivskins2022nanomechanical}.



\begin{figure*} 
	\centering
	\includegraphics[width=16cm]{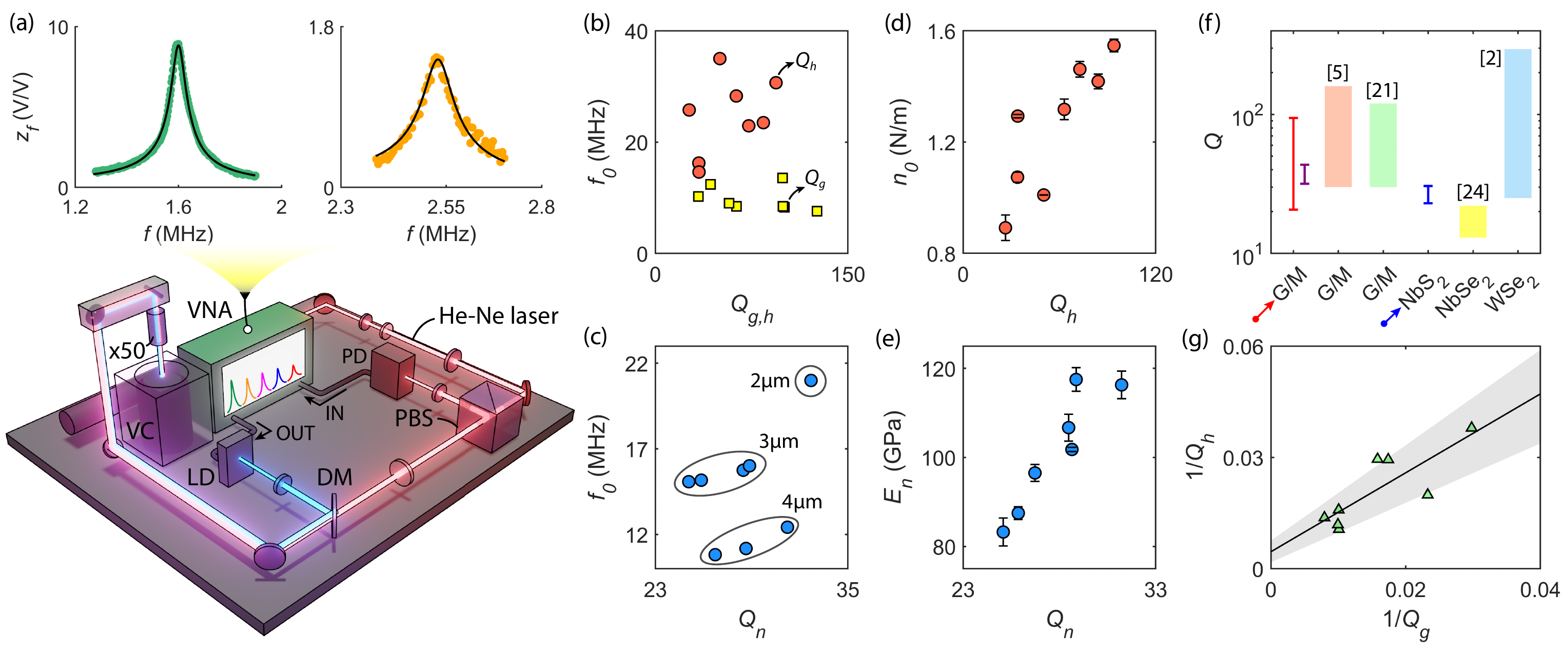}
	\caption{Dynamic characterization of the fabricated ALD nanomechanical resonators. (a) Top, the measured amplitude $z_f$ (points) of NbS$_2$ device D6, around the fundamental and second resonance, respectively; $z_f$ is fitted by a harmonic oscillator model (black lines) to extract resonance frequency and corresponding quality factor. Bottom, interferometry setup. ALD device is placed inside the vacuum chamber (VC). The diode laser (LD) is modulated by a vector network analyzer (VNA) to actuate the resonator, while intensity variations of the reflected He-Ne laser caused by resonator motion, are measured by photodiode (PD) and recorded with the VNA. PBS: polarized beam splitter; DM: dichroic mirror; x50: 50-fold objective. (b) The measured $Q$ versus $f_0$ for graphene resonators before ALD ($Q_g$, yellow points) and for heterostructure resonators after ALD ($Q_h$, red points). (c) The measured $Q_n$ versus $f_0$ for NbS$_2$ resonators. (d) $Q_h$ versus the extracted pretension $n_0$. (e) $Q_n$ versus the extracted Young's modulus $E_n$. (f) Comparison of $Q$ between this work (errorbars, marked with arrows in $x$-axis) and previous studies (bars). Purple errorbar shows the measured $Q$ of purely exfoliated heterostructure devices, as introduced in supplementary material. (g) The measured results of $Q_h^{-1}$ versus $Q_g^{-1}$ (green points) with a linear fit by Eq. \ref{Q} (black line). Gray shadow represents the fitting error bar.}   
	\label{fig: dynamics}
\end{figure*}

Let us now focus on the dynamics of the ALD resonators. We measured the dynamic response of the membranes with a laser interferometer \cite{vsivskins2022nanomechanical} (see Fig. \ref{fig: dynamics}a, bottom). A power modulated blue diode laser ($\lambda=$~\SI{405}{nm}) photothermally actuates the resonator, while the refection of a continuous-wave red He-Ne laser ($\lambda=$~\SI{632}{nm}) is sensitive to the time-dependent position of membrane. A vector network analyzer (VNA) provides a signal at drive frequency $f$ (OUT port) that modulates the blue laser intensity while the intensity of the red laser recorded by a fast photodiode is connected to the IN port. The VNA thus measures a signal $z_f$ that is proportional to the ratio of the membrane amplitude and actuation force. By sweeping the drive frequency $f$, we locate the resonance peak in the range from \SI{100}{kHz} to \SI{100}{MHz}. Laser intensities are set to \SI{0.3}{mW} (blue) and \SI{1.1}{mW} (red), respectively. These intensities are low enough for the resonator to vibrate in the linear regime. All measurements were performed at room temperature in vacuum at a pressure of $10^{-5}$ mbar. 

Figure~\ref{fig: dynamics}a (top inserts) shows the measured signal $z_f$ of NbS$_2$ device D6, at around the fundamental and second resonance frequency, respectively. By fitting $z_f$ to the response function of a harmonic oscillator, we extract $f_0=$~\SI{16.0}{MHz} with $Q=28.9$ and $f_1=$~\SI{25.3}{MHz} with $Q=34.4$. For vibrations of clamped drums, we can compute the resonance frequencies $f_i$ using \cite{lee2013high}
\begin{equation}
    \label{freq}
    f_i = \left( \frac{\mu_i}{2\pi}\right) \sqrt{\frac{D}{\sigma r^4} \left[ \mu_i^2+\frac{n_0 r^2}{D}   \right]}, i = 0,1,...,
\end{equation}
where $\sigma=\eta\rho t$ is the areal mass density, $\eta$ is a correction factor of mass considering the contaminations on resonators, and $\mu_i$ is a mode-specific factor. We have $\mu_1 = 2.4048$ for the fundamental mode and $\mu_2 = 3.8317$ for the second mode. For an ideal membrane, in which $n_0 r^2$ is much larger than the flexural rigidity $D$ and thus $n_0r^2/D\rightarrow\infty$, we have $f_1/f_0=\mu_1/\mu_0=1.59$; while for an ideal plate where $D$ is much larger than $n_0 r^2$, we have $n_0r^2/D\rightarrow0$ and thus $f_1/f_0=(\mu_1/\mu_0)^2=2.54$. The measured $f_1/f_0$ are $1.595\pm0.167$ and $1.959\pm0.642$ for heterostructure and NbS$_2$ devices, respectively (see supplementary tables~\ref{table: database1} and \ref{table: database2}), suggesting that the modes of heterostructure resonators are near the membrane limit, while the modes of NbS$_2$ resonators are in between the membrane and plate limit. We plot the extracted $Q$ versus $f_0$ for all heterostructure resonators ($Q_h$) and NbS$_2$ resonators ($Q_n$) in Figs.~\ref{fig: dynamics}b and \ref{fig: dynamics}c, respectively, including the quality factor $Q_g$ of the exfoliated graphene membranes before ALD. As expected from Eq.~\ref{freq}, $f_0$ decreases with increasing $r$ for the NbS$_2$ resonators, while $f_0^h$ for heterostructure resonators varies widely from 14.6 to \SI{30.7}{MHz}. This is attributed to the inhomogenieties like wrinkles and crumples in the heterostructures (see images in Fig.~\ref{fig:SI1}) and large differences in pretension. All measured resonance frequencies are comparable to those in literature reported for similar devices \cite{lee2013high,castellanos2012elastic}. 

The extracted values of $Q_n$ and $Q_h$ are also comparable to values of previously studied resonators made by exfoliation and CVD \cite{aguila2022fabry,zhu2022frequency,ye2017atomic,ye2021ultrawide}, as illustrated in Fig. \ref{fig: dynamics}f. To gain insight into the damping, we plot $Q_h$ versus $n_0$ for heterostructure resonators and $Q_n$ versus $E_n$ for NbS$_2$ resonators, respectively, as plotted in Figs.~\ref{fig: dynamics}d and \ref{fig: dynamics}e. For both cases, we observe a linear relation, indicating that pretension plays a more important role on damping than bending rigidity for heterostructure resonators, while it is on the other way around for NbS$_2$ resonators. This is exactly as expected based on the ratio $f_1/f_0$. On the other hand, we do not see clear relations of $Q_h$ versus $E_h$ and $Q_n$ versus $n_0$ as plotted in Fig. S2a and S2b, respectively. 

Concerning the effective masses, we determine the correction factors $\eta_h$ and $\eta_n$ by substituting the measured $f_0$, and the extracted $n_0$ and $E$ into Eq.~\ref{freq} (see values in supplementary tables~\ref{table: database1} and \ref{table: database2}). We obtain $\eta_h=1.34\pm0.92$ and $\eta_n=2.87\pm0.83$, respectively. The high $\eta_n$ of NbS$_2$ devices is attributed to the contaminations from the PDMS stamping. The values $\eta_h$ for the heterostructure are surprisingly close or even below 1. This suggests the absence of any residues and possibly even the thinning of the graphene membrane during the ALD process, while the ALD MoS$_2$ layer is mainly deposited on top of the suspended graphene membrane instead of bottom (see supplementary S2). 


We also observe a general decrease of quality factor in heterostructure resonators after ALD ($Q_h<Q_g$), as shown in Fig.~\ref{fig: dynamics}b. Considering the dissipation mechanism for two parallel membranes, the overall $Q_h$ can be modeled as 
\begin{equation}
    \label{Q}
    {1}/{Q_h} = {\alpha}/{Q_g}+{1}/{Q_{m}},
\end{equation}
where $\alpha$ can be different than 1 on account of structural changes in the graphene because of the ALD process, and $Q_{m}^{-1}$ is a fit parameter that represents the damping in the heterostructure originating from the ALD MoS$_2$. We fit the measured $Q_h^{-1}$ versus $Q_g^{-1}$ with Eq.~\ref{Q} (see Fig.~\ref{fig: dynamics}g) and extract $\alpha=1.1\pm0.1$ and $1/Q_{m}=4.7\pm3.1\times 10^{-3}$. The fact that the obtained $\alpha$ (within errors) is close to 1, provides evidence that there little to none increase of the dissipation in the graphene during the ALD process. A control experiment has been done with purely exfoliated graphene/MoS$_2$ heterostructures (see Fig.~\ref{fig:SI}), giving us $\alpha=1.1\pm0.2$ and $1/Q_{m}=17.6\times 10^{-3}$. The lower $1/Q_{m}$ of ALD heterostructure compared to exfoliated layers can be attributed to a better conformality of the ALD layer and the absence of contamination by transfer polymers.

Although ALD is known to be capable of wafer-scale synthesis, the dimensions of our fabricated devices are still quite small due to the use of exfoliation in the fabricating process. A strategy could be to grow transferless suspended CVD 2D material membranes like graphene \cite{pezone2022sensitive}, and subsequently grow ALD material heterostructures from them. In addition, ALD could benefit from a method to precisely control the flatness, so as to avoid the cragged surfaces of nanoscale devices as illustrated in Figs.~\ref{fig:SI1}a and \ref{fig:SI1}b.       

In conclusion, we presented the fabrication and mechanical characterization of nanomechancial resonators consisting of ALD 2D materials. We developed two PEALD based approaches to suspend ALD flakes on a patterned Si/SiO$_2$ substrate: one is dry transfer using PDMS (exfoliate ALD Nb$_2$ flakes from glassy carbon); the other is ALD deposition of MoS$_2$ on mechanically exfoliated suspended graphene drums. AFM indentation allows us to determine their Young's moduli as $101.4\pm$ \SI{13.3}{GPa} and $951.7\pm$ \SI{161.0}{GPa}. Using an optomechanical method, we extracted their resonance frequencies and the corresponding quality factors. All of the above parameters are well comparable to the reported values of exfoliated and CVD resonators. We found experimental indications that the dissipation of ALD MoS$_2$ membranes in heterostructures is roughly 3.7 times lower than that of purely exfoliated MoS$_2$ membranes, which is promising for high-performance 2D heterostructure resonators. Our results show possibilities toward exploiting ALD technique for nanomechancial resonators in future explorations on atomically thin tunable resonators and 2D sensors. Meanwhile, the thickness-controllable ALD heterostructures could provide valuable insight in interactions at 2D interfaces, which can bring significant improvements in device performance and lead to new functionalities.

\textbf{Conflict interest} The authors declare that they have no competing financial interests.

\textbf{Data Availability} The data that support the findings of this study are available from the corresponding authors upon reasonable request.


\bibliography{main}
\pagebreak
\onecolumngrid

\section*{Supplementary Materials}

\subsection*{S1: AFM indentation measurements on fabricated ALD resonators} 

We fit the measured curves of $F$ versus $\delta$ to extract the pretension $n_0$ and Young's modulus $E$ of the fabricated resonators. The applied force $F$ equals the product of the cantilever stiffness $k_c$ and its deflection $\Delta z_c$. We use a cantilever with $k_c=$ 53.7 $\pm$ \SI{0.1}{N/m}, and repeat the indentation measurement three times for each device.

The classical relation for the bending rigidity, $D=Et^3/(12(1-\nu^2))$, in general, is not valid for multilayer 2D materials, where the interlayer shear interactions are weak and slippage is inevitable. As a result, a calibration factor $\gamma$ is induced to describe this interaction, giving the formula as $D=\gamma Et^3/(12(1-\nu^2))$. Since the layers number of graphene and MoS$_2$ in the fabricated heterostructures are 40 and 13 roughly, we adopt $\gamma_g=0.1$ and $\gamma_m=0.4$ from literature \cite{wang2019bending}, respectively. We also assume $\gamma_n=\gamma_m$ due to their similar lattice structures. For NbS$_2$ resonators, we can directly fit the measured $F$ versus $\delta$ to Eq. 1 to obtain $n_0$ and $E_n$. However, for heterostructure resonators, considering the different mechanical properties of graphene and MoS$_2$ layers, their effective Young's modulus $E_h$ and effective bending rigidity $D_h$ are given by \cite{ye2017atomic,vsivskins2022nanomechanical}
\setcounter{equation}{0}
\begin{equation}
    \label{eq:S1}
    \begin{array}{l}
   E_ht_h = E_gt_g+E_mt_m\quad \text{and}\quad D_ht_h = D_gt_g+D_mt_m,
   \end{array}
    \tag{S1}
\end{equation}
respectively, where $t_h=t_m+t_g$. As a result, the relation of $F$ versus $\delta$ for heterostructure resonators is expressed as
\setcounter{equation}{1}
\begin{equation}
    \label{eq:S2}
    F = \left[\frac{4\pi}{3r^2}\cdot \left(\frac{\gamma_g E_gt_g^4}{1-\nu_g^2}\cdot \frac{1}{t_h}+\frac{\gamma_m E_mt_m^4}{1-\nu_m^2}\cdot \frac{1}{t_h}\right) \right]\delta+n_0\pi\delta+(E_gt_g+E_mt_m)q^3 \left(\frac{\delta^3}{r^2}\right).
    \tag{S2}
\end{equation}
Using the values of $t_g=13.3$ \SI{}{nm}, $t_m=7.8$ \SI{}{nm}, $\nu_g=0.165$, $\nu_m=0.25$, $\gamma_g=0.1$ and $\gamma_m=0.4$, the part inside parenthesis of the first term in Eq. \ref{eq:S2} can be rewritten as $(E_gt_g\cdot11.46+E_mt_m\cdot9.60)\times10^{-18}$. This part is then replaced by $E_ht_h\cdot9.60\times10^{-18}$ and $E_ht_h\cdot11.46\times10^{-18}$ separately in the fit, causing only a deviation $<0.7\%$ to the extracted $n_0$. 

\subsection*{S2: Experimental results of nanomechanical characteristics for all fabricated ALD resonators in this work} 

TABLE \ref{table: database1} gives the measured parameters of all ALD heterostructure devices, including radius $r$ of drums, Young's modulus $E_h$, pretension $n_0$, fundamental resonance frequency $f_0^h$ and $f_0^g$ (corresponding to the graphene membrane before ALD), quality factor $Q_h$ and $Q_g$, modes ratio $f_1^h/f_0^h$, and the calibration factor $\eta$ with respect to the mass of membrane. TABLE \ref{table: database2} gives the measured parameters of all ALD NbS$_2$ devices. We cannot extract a precise value of $E_n$ for device D8 due to its small size, which needs a quite large loading $F$ to achieve the effective indentation with cubic regime. The second resonance frequency $f_1$ of device D8 is missed as well, since we use a low pass filter (up to \SI{60}{MHz}) on VNA during the dynamic measurements. 

Note that in the calculation of $\eta$ for heterostructure resonators, we have already used the relation $\sigma =\eta(\rho_g t_g + \rho_m t_m)$, where the thickness $t_m$ of ALD MoS$_2$ is determined by scanning the scratch of MoS$_2$ layer on Si/Si$_2$ substrate. We assume that ALD MoS$_2$ mainly grows on top of the graphene membrane, since the average value of $\eta$ in TABLE~\ref{table: database1} is close to 1, indicating that the thickness of ALD MoS$_2$ layer in heterostructure is roughly equal to that on substrate. For heterostructure devices D4 and D5, $\eta$ is larger than 1, which might result from a small quantity of ALD MoS$_2$ deposited on the bottom of graphene.    

Figure~\ref{fig:SI1} shows the AFM scanning images of ALD NbS$_2$ device D1 and D2, as well as all ALD heterostructure devices. We can observe the visible polymer residues, crumples and wrinkles on these devices, which significantly affect their static and dynamic properties. More details about the TEM images of ALD MoS$_2$ can be found in our previous work \cite{sharma2020large}.

\begin{table} [h]
	\centering
	\label{table: database1}
	\caption{Nanomechanical properties of graphene/MoS$_2$(ALD) heterostructure resonators.}
	\begin{tabular}{lllllllll}
		\hline 
		Device & $E$ (\SI{}{GPa}) & $n_0$ (\SI{}{N/m}) & $f^g_0$ (\SI{}{MHz}) & $Q_g$ & $f^h_0$ (\SI{}{MHz}) & $Q_h$ & $f^h_1/f^h_0$ & $\eta$ \\ 
		\hline 
		D1 & 951.7 & 1.418 & 8.3 & 100.6 & 23.5 & 84.2 & 1.41 & 1.16 \\ 
    	\hline 
		D2 & 771.4 & 1.547 & 13.6 & 99.1 & 30.7 & 94.0 & 1.49 & 0.61 \\ 
			\hline 
    	D4 & 825.5 & 1.074 & 8.5 & 63.2 & 16.2 & 33.9 & 1.58 & 2.03 \\ 
			\hline 
		D5 & 1095.7 & 1.293 & 9.0 & 57.5 & 14.6 & 34.0 & 1.71 & 3.23 \\ 
			\hline 
		D6 & 1182.6 & 1.462 & 7.6 & 125.9 & 22.9 & 72.8 & 1.53 & 1.44 \\ 
			\hline 
		D7 & 950.7 & 1.318 & 8.4 & 99.4 & 28.3 & 63.0 & 1.67 & 0.78 \\ 
			\hline 
		D8 & 1162.9 & 0.892 & 10.2 & 33.6 & 25.8 & 26.4 & 1.92 & 1.01 \\ 
			\hline 
		D10 & 820.9 & 1.010 & 12.4 & 43.0 & 35.0 & 50.3 & 1.45 & 0.43 \\ 
		\hline  
	\end{tabular} 
\end{table}

\begin{table} [h]
	\centering
	\label{table: database2}
	\caption{Nanomechanical properties of ALD NbS$_2$ resonators.}
	\begin{tabular}{lllllllll}
		\hline 
		Device & $r$ (\SI{}{\micro\meter}) & $E_n$ (\SI{}{GPa}) & $n_0$ (\SI{}{N/m}) & $f_0$ (\SI{}{MHz}) & $Q_n$ & $f_1/f_0$ & $\eta$ \\ 
		\hline 
		D1 & 4 & 116.3 & 0.899 & 12.4 & 31.2 & 1.81 & 2.05\\ 
    	\hline 
		D2 & 4 & 101.8 & 0.447 & 11.2 & 28.6 & 2.26 & 2.12\\ 
			\hline 
    	D3 & 4 & 96.5 & 0.593 & 10.8 & 26.7 & 1.58 & 2.20\\ 
			\hline 
		D4 & 3 & 87.5 & 1.320 & 15.2 & 25.9 & 1.67 & 3.30\\ 
			\hline 
		D5 & 3 & 106.7 & 1.077 & 15.8 & 28.5 & 1.72 & 3.60\\ 
			\hline 
		D6 & 3 & 117.5 & 1.005 & 16.0 & 28.9 & 1.57 & 3.79\\ 
			\hline 
		D7 & 3 & 83.2 & 0.797 & 15.1 & 25.1 & 3.10 & 3.05\\ 
			\hline 
		D8 & 2 & $-$ & $-$ & 21.0 & 32.7 & $-$ & $-$ \\ 
		\hline  
	\end{tabular} 
\end{table}

\setcounter{figure}{0}
\begin{figure}[H]
   \centering
   \renewcommand{\thefigure}{S\arabic{figure}}
    \includegraphics[width=13cm]{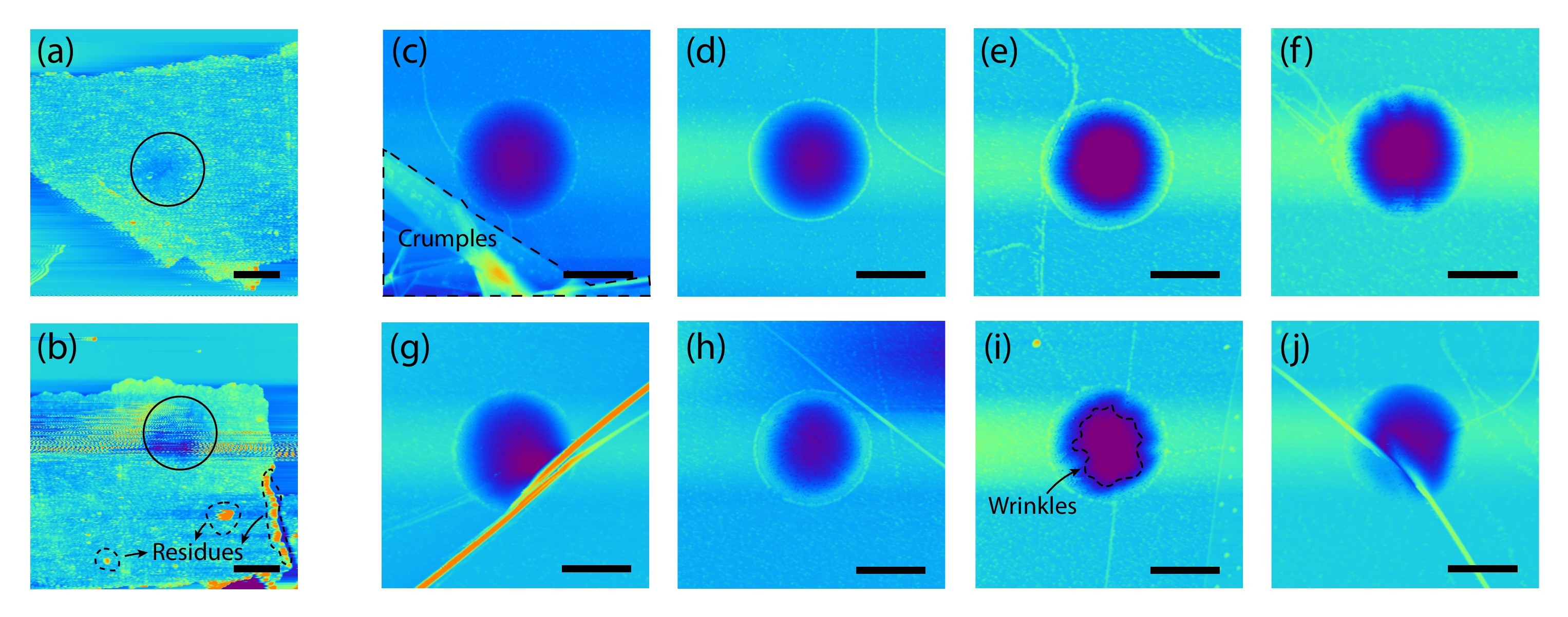}
    \caption{AFM scanning results of our fabricated ALD devices. (a) and (b) ALD NbS$_2$ device D1 and D2, respectively. (c)$-$(j) All measured ALD heterostructure devices D1 to D10, except the broken ones D3 and D9. Scale bar is \SI{5}{\micro\meter}.}
    \label{fig:SI1}
\end{figure}

Figures~\ref{fig:SI2}a and \ref{fig:SI2}b give the obtained $E_h$ versus $Q_h$ for heterostructure resonators and $n_0$ versus $Q_n$ for NbS$_2$ resonators, respectively. Unlike the proportional relations shown in Figs.~\ref{fig: dynamics}b and \ref{fig: dynamics}c in the main text, we see $E_h$ versus $Q_h$ and $n_0$ versus $Q_n$ are irregular here.

\setcounter{figure}{1}
\begin{figure}[H]
   \centering
   \renewcommand{\thefigure}{S\arabic{figure}}
    \includegraphics[width=10cm]{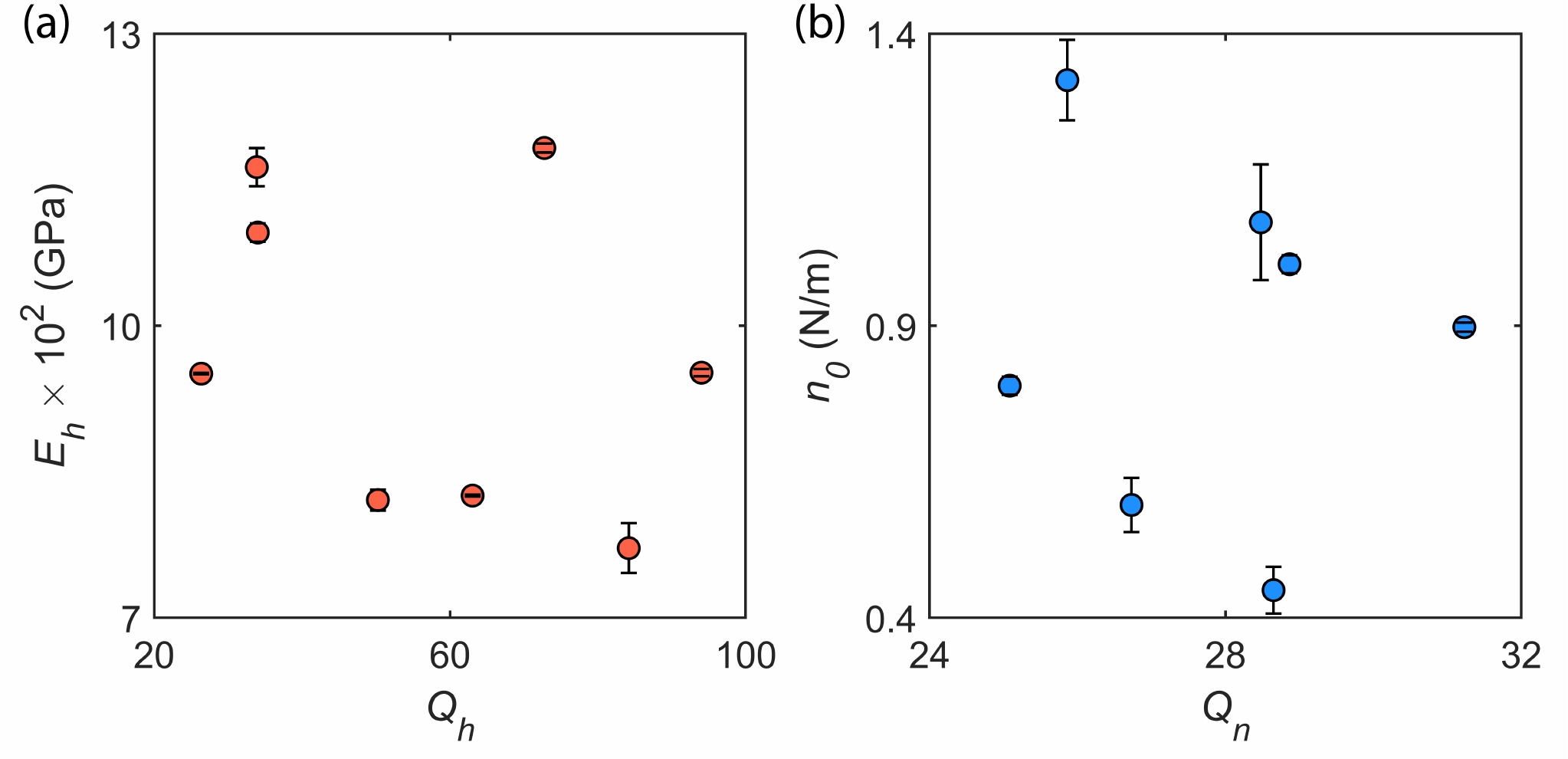}
    \caption{Discussion on mechanical properties of our fabricated ALD devices. (a) Effective Young's modulus $E_h$ versus quality factor $Q_h$ for heterostructure resonators. (b) Pretension $n_0$ versus quality factor $Q_n$ for NbS$_2$ resonators.}
    \label{fig:SI2}
\end{figure}

\subsection*{S3: Experimental results of quality factors for purely exfoliated graphene/MoS$_2$ heterostructure resonators} 

To shed light on the energy dissipation $Q_{m}^{-1}$ of MoS$_2$ layer in the heterostructure, purely exfoliated graphene/MoS$_2$ heterostructure devices are fabricated (Fig.~\ref{fig:SI}a) and measured in interferometry setup. As plotted in Fig.~\ref{fig:SI}b, the measured $Q_h^{-1}$ versus $Q_g^{-1}$ is fitted with Eq. \ref{Q} and thus extract $\alpha=1.1\pm0.2$ and $1/Q_{m}=17.6\pm1.9\times 10^{-3}$.

\setcounter{figure}{2}
\begin{figure}[H]
   \centering
   \renewcommand{\thefigure}{S\arabic{figure}}
    \includegraphics[width=10cm]{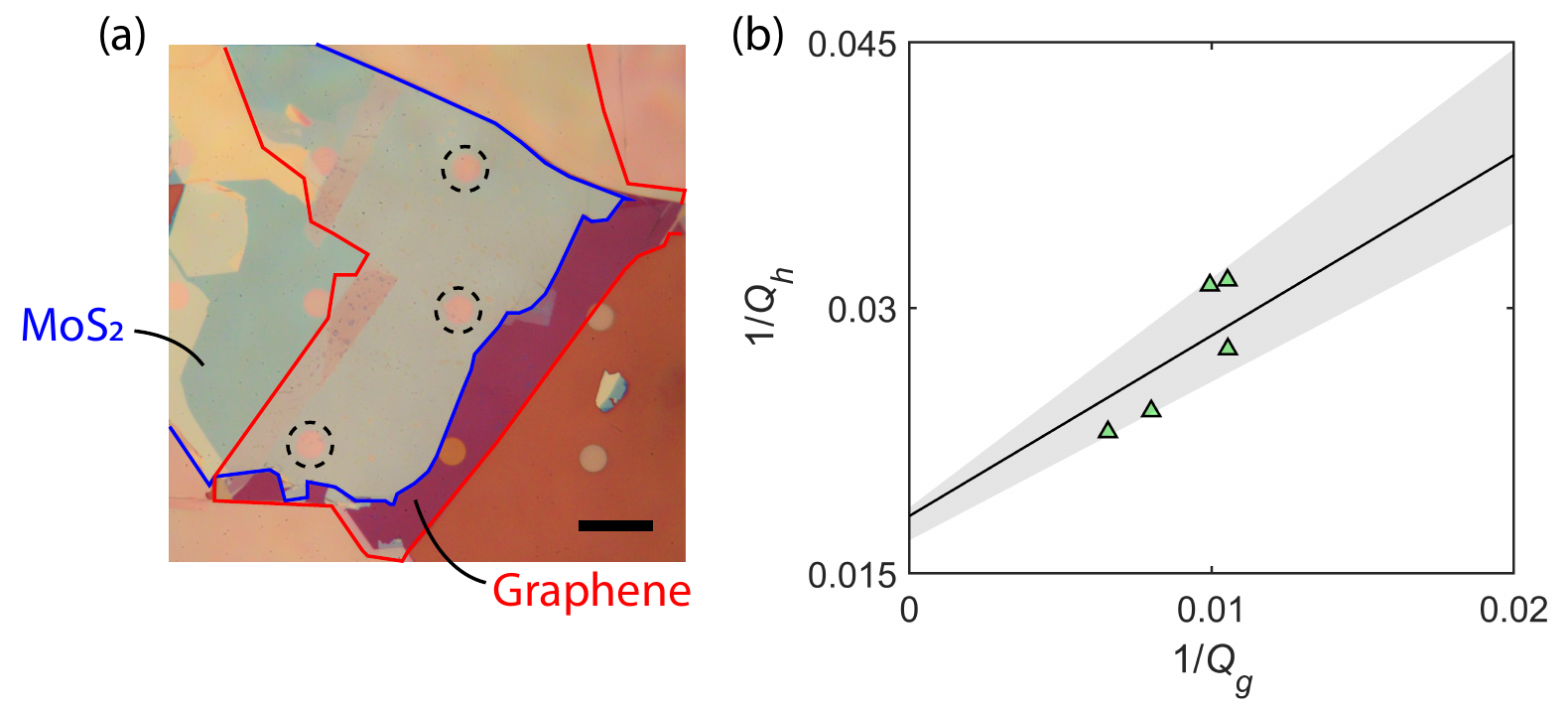}
    \caption{Discussion on the dissipation of MoS$_2$ layer in purely exfoliated graphene/MoS$_2$ heterostructure resonators. (a) Optical images of partial fabricated devices (marked by dotted circles), where the red and blue frames represent graphene (bottom) and MoS$_2$ (top) flakes, respectively. Scale bar is \SI{20}{\micro\meter}. (b) The measured results of $Q_h^{-1}$ versus $Q_g^{-1}$ (green points) and its fitting with Eq. \ref{Q} in the main text (black line and shadow).}
    \label{fig:SI}
\end{figure}

\end{document}